# Valley-spin polarized Landau levels in a monolayer semiconductor

Zefang Wang, Jie Shan*, and Kin Fai Mak*

Department of Physics and Center for 2-Dimensional and Layered Materials

The Pennsylvania State University, University Park, Pennsylvania 16802-6300, USA

*Correspondence to: jus59@psu.edu; kzm11@psu.edu

**Electrons in monolayer transition metal dichalcogenides (TMDs) possess both the valley and spin degree of freedom [1]. These internal quantum degrees of freedom have provided an ideal laboratory for exploring both new physical phenomena and electronics and photonics applications [1,2]. Valley- and spin-dependent optical and electrical properties, originated from Berry curvature effects, have been recently demonstrated [3-8]. Such Berry curvature effects, together with strong spin-orbit interactions, can further generate unconventional Landau levels (LLs) under a perpendicular magnetic field that support valley- and spin-polarized chiral edge states in the quantum Hall regime [9,10]. The unique LL structure, however, has not been demonstrated in TMDs. Here we report the observation of fully valley- and spin-polarized LLs in high-quality $WSe_2$ monolayers achieved using a van der Waals heterostructure device platform [11]. Handedness-resolved optical reflection spectroscopy has been applied to probe the inter-LL transitions at individual valleys. The LL structure has been constructed. A doping-induced mass renormalization driven by the strong Coulomb interactions has also been observed. Our results open the door for studies of the unconventional LL physics and quantum Hall effect in monolayer TMDs.**

Monolayer transition metal dichalcogenides (TMDs) ($MoS_2$, $MoSe_2$, $WS_2$, $WSe_2$ etc.) with a honeycomb lattice structure are multi-valley direct bandgap semiconductors [1,2]. Dirac electrons in these materials are characterized by both the valley (K and K') and the spin indices [1]. Under a perpendicular magnetic field, they have been predicted to form discrete Landau levels (LLs) [9,10] that are distinct from the case of two-dimensional (2D) electrons both in conventional semiconductor quantum wells (QWs) [12-14] and in graphene [15]. Because of the broken sublattice symmetry and of the valley-contrasting $\pi$ Berry's phase associated with the conduction and valence bands in monolayer TMDs [9,10,15,16], the zeroth LLs at the K and K' valleys are split by the material's band gap and are thus valley-polarized. Specifically, they are fixed to the conduction band minimum and the valence band maximum of the K and K' valley, respectively, if Zeeman effect is ignored. Valley-polarized LLs have been demonstrated in non-centrosymmetric QWs [13,14] and in gapped bilayer graphene [17-19]. There are, however, important differences between these systems and the TMDs. In QWs, the Berry curvature effects do not play an important role. Instead, application of a symmetry-lowering strain is required to produce valley-polarized LLs [13]. On the other hand, although the Berry curvature effects are responsible for lifting the valley degeneracy in both TMDs and gapped graphene, the strong spin-orbit interactions (SOIs), present only in TMDs, further spin-polarize the LLs at each valley [9,10]. Compared to the LLs in graphene with spontaneous SU(4) symmetry breaking [20,21], the fully lifted LL degeneracy in TMDs, applicable to *all* LLs, is more robust. To date, the unique LL structure in monolayer TMDs has not been observed.

A major challenge in observing the unique LL structure in monolayer TMDs lies in the sample quality. In comparison to conventional III-V QWs [12], the electron cyclotron energy $\hbar\omega_c \approx \frac{\hbar e}{m_e}B$ in monolayer TMDs is nearly an order of magnitude smaller under the same magnetic field $B$ because of their large electron band masses $m_e$ (here $\hbar$ is Planck's constant and $e$ is the elementary charge) [9, 10]. As we show below, monolayer TMDs with mobilities of ~ 1000 cm$^2$/Vs or higher are required to observe the LLs at low temperature under a laboratory scale magnetic field. Moreover, the study of the LLs by optical measurements are hampered by the strong excitonic effects that are characteristic of monolayer TMDs [2, 22]. Although Shubnikov-de Haas oscillations (SdHO) in the dc conductivity has been observed in monolayer TMDs [11, 23], the valley-spin polarized LL structure cannot be directly probed by transport measurements.

In this work, we investigate the LLs in high-quality WSe$_2$ monolayers by handedness-resolved optical reflection spectroscopy at 3 K. We have employed the van der Waals heterostructure device platform to fabricate dual-gate transistors of monolayer WSe$_2$ (Ref [11]) (See Methods for details). The WSe$_2$ samples are encapsulated by thin films of hexagonal boron nitride (hBN) with few-layer graphene as contact and gate electrodes (Fig. 1a). The combination of hBN encapsulation that minimizes disorder and the dual local gates that enable tuning of the doping density to significantly suppress the excitonic effects [22] has allowed the observation of inter-LL transitions in monolayer WSe$_2$ under a perpendicular magnetic field as low as 3 Tesla. We have decomposed the optical reflectance contrast into responses to the right ($r$) and left ($l$) circularly polarized light to probe the optical transitions at the K and K' valley independently. The valley contrasting optical selection rules have been the basis for optical control of the valley polarization in monolayer TMDs under $B = 0$ [1, 3-5] and have been shown to remain valid even under finite $B$-fields [24-30]. In particular, the dipole allowed transitions are $-(n+1) \leftrightarrow n$ at the K valley (for $r$ polarization) and $-n \leftrightarrow n+1$ at the K' valley (for $l$ polarization), where $n = 0, 1, 2…$ is the LL index [24, 25].

The dual-gate structure allows us to independently tune the vertical electric field and the doping density in monolayer WSe$_2$. Here we focus on the doping effects under zero electric field (See Supplementary Section 1 for details on gating and basic optical and electrical characterization of the devices). Figure 1b shows the unpolarized reflectance contrast $\frac{\Delta R}{R}$ as a function of doping density $\sigma$ and photon energy $E$ (1.64 – 1.75 eV) under $B = 0$. In this region, the spectrum is dominated by the A exciton absorption around 1.73 eV (regime I) and charged excitons around 1.70 eV under relatively low electron or hole doping densities (regime II), as previously reported [1]. When $\sigma > \sigma_0 = 7.0 \times 10^{12}$ cm$^{-2}$ (regime III), a kink is observed, corresponding to doping into the spin-split upper conduction band c1 as shown in the electronic structure [31] (Fig. 1c). (For more evidence see below and in Supplementary Section 1.2.) Around the K and K' points of the Brillouin zone, the low-energy electronic structure consists of spin-split conduction bands (c1, c2) and valence bands (v1, v2) with a splitting of ~ 30 and 400 meV by the SOIs, respectively [2, 31]. Optical transitions are allowed only between bands of the same spin with the lowest energy transitions (v1 $\leftrightarrow$ c1) giving rise to the A exciton [2]. The observed kink and a subsequent small blue shift of the absorption edge (Supplementary Fig. S4) in regime III can be understood as the emergence of Pauli blocking of optical transitions to the Bloch states in band c1. Because of the Schottky nature of the contacts of our devices, calibration of the doping density $\sigma$ by the Hall effect was not possible. Instead, $\sigma$ was calibrated from the applied gate voltages using capacitances determined from the dielectric constant ($\varepsilon \approx 2.7$) and thickness (~ 20 nm on both sides) of the hBN thin films. We took $\sigma = 0$ (i.e. Fermi level at the band edge of c2) where the neutral exciton feature just disappears. The (systematic) error introduced in $\sigma$ is estimated to be ~10%. (See Supplementary Section 1.4 for the measurement of $\varepsilon$ and discussion on the error analysis.) Below we will study the *B*-field effects focusing on the region given by the dashed box, where the excitonic effects are significantly reduced by doping and optical transitions remain sharp.

Figure 2b & 2c show the reflectance contrast $\frac{\Delta R}{R}$ under 9 T for the $r$ (K) and $l$ (K') channel, respectively. Unlike the case of $B = 0$, the $l$ and $r$ spectra under 9 T are distinct, a manifestation of broken time-reversal symmetry (See Supplementary Section 2 for extra data). The optical contrast, equivalent to the optical conductivity, oscillates with doping density. (Representative vertical cuts at photon energy $E = 1.665$ eV for the two valleys are shown in Fig. 2a. The corresponding optical ellipticity $e = r - l$, reflecting the sample's variable magnetic moment, is shown in Fig. 2d.) This is analogous to the SdHO in the dc conductivity [11, 23], a characteristic signature of LLs in transport. On the other hand, for a given doping density (a horizontal cut), multiple fully resolved transitions with equal energy spacing can be identified. The energy, and to a lesser extent the spacing (see below), of these transitions are doping dependent. We tentatively assign them inter-LL transitions. These transitions disappear sequentially with increasing doping as a result of Pauli blocking of the LLs of band c1. We trace the critical doping densities $\sigma_n^*$ required to fill the $n^{\text{th}}$ LL of band c1 from the onset of Pauli blocking of the corresponding inter-LL transition (marked by horizontal dashed lines in Fig. 2b & 2c). Note that such densities are distinct for the K and K' valley, a clear indication of LLs being valley polarized. We summarize the critical densities for the first several LLs of band c1 at varying magnetic fields in Fig. 3. In order of density increase, the LLs are labeled as $n_K = 0, 1, 2, \ldots$ for the K valley, and $n_{K'} = 1, 2, \ldots$ for the K' valley [9, 10, 24, 25]. We note that $\sigma_{n_K=0}^*$ (equivalent to the energy of the $n_K = 0$ LL) decreases linearly with *B*-field although the zeroth LLs have been predicted to be non-dispersive in *B*-field from the Berry curvature effects [9, 10]. This magnetic field dependence arises from the Zeeman effect [24-30] and will be used to determine the electron effective magnetic moment below.

Next we determine the LL degeneracy. When the Fermi level is shifted from the (*n*-1)$^{\text{th}}$ to the $n^{\text{th}}$ LL in the K or the K' valley, the LL degeneracy $f$ can be evaluated as: $\sigma_n^* - \sigma_{n-1}^* = f \frac{B}{\Phi_0}$, where $\Phi_0 = 4.14 \times 10^{-15}$ Tm$^2$ is the magnetic flux quantum [12]. From the slopes of Fig. 3 we obtain $f = 3.4 \approx 3$ for $n_K = 1$ and 2, and $f = 4.4 \approx 4$ for $n_{K'} = 2, 3, 4$. The 10% overestimate in $f$ is most likely due to the systematic error in $\sigma$ as mentioned above. The unity change from $f = 3.4$ to $f = 4.4$, together with the oscillatory magnetic moment as a function of doping density (Fig. 2d), provide strong evidence that the degeneracy of each LL is 1, i.e. *each LL is fully valley and spin polarized*. In this picture, the first three LLs of band c1 in the K valley ($n_K = 0, 1, 2$) are below all the other LLs of band c1 in energy. To fill the $n_K = 1$ (or 2) LL after filling $n_K = 0$ (or 1), one needs to fill three LLs (thus $f = 3$) including 1 LL of band c1 at the K valley and two LLs of band c2 (one from each valley). Similarly, to fill successively each of the higher energy LLs, one needs to fill four LLs (thus $f = 4$) including one LL of the K valley and one LL of the K' valley of band c1, and two LLs of band c2. In comparison, the (*non-spin* *resolved*) transport measurements have determined a LL degeneracy of 2 in hole-doped monolayer WSe$_2$ (Ref [23]), which is in agreement with our result, and a LL degeneracy between 2 and 4 in electron-doped monolayer MoS$_2$ (Ref [11]) from the SdHO in the dc conductivity. The discrepancy between the two TMD compounds could be due to the small spin-orbit splitting in the conduction bands of MoS$_2$.

We now perform a more careful analysis of the inter-LL transition spectrum. As an example, we choose doping density $\sigma = 5.9 \times 10^{12}$ cm$^{-2}$. Other doping results are similar. In Fig. 4b, we show several representative reflectance contrast spectra under 3 - 9 T for both the *r*(K) and *l*(K') channel. Resonances are visible under a field as low as 3 T. Under 9 T, the peaks are separated by ~ 4 meV and the peak width is ~ 2 meV (corresponding to a relaxation time of ~ 3x10$^{-13}$ s and a mobility of ~ 1500 cm$^2$/Vs if an effective mass of $0.4m_0$ is assumed [31]). We label these inter-LL transitions in order of energy increase as $N^+$ and $N^-$ (= 1, 2,…) for the *l*(K') and *r*(K) channel, respectively. The transition energies $E_{N^\pm}$ as a function of *B*-field are summarized in Supplementary Fig. S9. These energies can be described as the sum of the LL energies and the Zeeman energies [24, 25]

$$E_{N^\pm} \approx E_0 + \frac{e\hbar}{m_{e/h}}B + (N^\pm - 1)\frac{e\hbar}{m_r}B \pm (\mu_e - \mu_h)B, \qquad (1)$$

where the "+" sign and conduction band mass $m_e$ in the second term are for the K' valley; and the "-" sign and the valence band mass $m_h$ in the second term are for the K valley. In Eqn. (1) the first term $E_0$ is the transition energy under the given doping density for $B = 0$. The second and third term are the LL energies of the conduction and valence band with consideration of the optical selection rules; $m_r^{-1} = m_e^{-1} + m_h^{-1}$ is the electron-hole reduced mass. The fourth term is the Zeeman energies with $\mu_e$ and $\mu_h$ denoting the effective electron and hole magnetic moment, respectively. (See Supplementary Section 2.3 for more detailed derivations.) To isolate the LL energies, we plot our experimental result $\frac{E_{N^+}+E_{N^-}}{2} = E_0 + (2N-1)\frac{e\hbar}{2m_r}B$ as a function of $B^* = (2N-1)B$ in Fig. 4c. Excellent agreement is obtained with $E_0 = 1.664$ eV and $m_r \approx 0.268 \pm 0.003\, m_0$ ($m_0$ is the free electron mass). The latter is higher than $0.2m_0$ determined from the first principles ($m_e \approx m_h \approx 0.4m_0$ [31]). (See below for more discussions on the doping-induced mass renormalization.) In the inset of Fig. 4c we show the *B*-dependence of $|E_{N^+} - E_{N^-}|/2$. If the small mass difference between band v1 and c1 is ignored, the slope yields $|\mu_e - \mu_h| = 0.12 \pm 0.02$ meVT$^{-1}$, which is also in good agreement with previous magneto-luminescence studies [28-30].

Finally we construct the unique LL structure of monolayer WSe$_2$ in Fig. 4a. For simplicity, we neglect the small band mass difference for band v1, c1 and c2 [31] without assuming any specific band mass values. This allows us to determine the LL energies in band c1 from the inter-LL transition energies. We take the Fermi energy $E_F = 0$ at the edge of the spin-split lower conduction band c2 (where electron doping density $\sigma = 0$). We note that successive Pauli blocking of inter-LL transitions in the K valley occurs in the regime of $\sigma > \sigma^*_{n_K=0}$ when the Fermi level sweeps through each LL of band c1 in that valley. The corresponding change in $E_F$ between successive filling (the conduction band cyclotron energy) is half of the energy spacing between the successive inter-LL transitions, i.e. $\hbar\omega_c \approx 2$ meV at 9 T. The conversion rate of $\sigma$ to $E_F$ can thus be determined. Below the critical density $\sigma^*_{n_K=0}$, however, doping only occurs in band c2 and $E_F$ varies with $\sigma$ at twice the rate since the number of electron species is halved. The resultant energy $E_F$ is shown on the right axis of Fig. 2. Equipped with the $\sigma$ to $E_F$ conversion relationship, we can also determine the electron effective magnetic moment to be $\mu_e = 0.13 \pm 0.02$ meVT$^{-1}$ from the slope of the $B$-field dependence of the $n_K = 0$ LL energy (a pure Zeeman effect, Fig. 3). Combined with the extracted value for $|\mu_e - \mu_h|$, the hole effective magnetic moment is estimated to be $\mu_h = 0.25 \pm 0.02$ meVT$^{-1}$. Knowing the Zeeman shift and the energy of the $n_K = 0$ LL in band c1, we can now fix the band edge of band c1. This yields a spin-orbit splitting of 29 meV for the conduction bands, which is also in good agreement with first principles calculations [31]. Using the electron effective magnetic moment of opposite sign for band c2, we can also determine the LLs for band c2. And using the inter-LL transition energies under a known doping density (for instance, $5.9 \times 10^{12}$ cm$^{-2}$) we determine the LLs of band v1. The complete LL structure is shown in Fig. 4a. (See Supplementary Section 3 for more details.)

So far we have ignored interaction effects in the discussion. We note that first, the excitonic effects are known to be very strong in monolayer TMD semiconductors at relatively low doping densities [2]. In this limit, the cyclotron gyroradius is much larger than the exciton Bohr radius and the effect of the $B$-field is a weak diamagnetic correction $\propto B^2$ (ref [22, 32]). When the gyroradius becomes comparable to or smaller than the Bohr radius for weak excitonic binding or under high fields, the field dependence crosses over from $\propto B^2$ to $\propto B$, and the excitonic transitions become band-to-band like inter-LL transitions [22, 32]. Our experiment belongs to the latter case. Second, we show in Fig. 4d the doping dependence of the reduced mass $m_r$ extracted from the magnetic-field dependence of the inter-LL transition energies (See Supplementary Section 4 for details of the analysis). With increasing doping density, $m_r$ decreases, goes through a step rise at $\sigma_0$ (dashed line), and then continues to decrease. The phenomenon of doping-induced mass renormalization has been studied in other 2D electron systems [33]. The interaction strength is measured by the Wigner-Seitz radius, $r_s = \frac{g_s g_v}{\sqrt{\pi \sigma} a_B^*}$, where $g_s$ and $g_v$ denote the spin and valley degeneracy, respectively, and $a_B^* = a_B \varepsilon / (\frac{m_{e0}}{m_0})$ is the Bohr radius $a_B$ corrected by the dielectric constant $\varepsilon$ of the medium (hBN) and the bare conduction band mass $m_{e0} \approx 0.4 m_0$ [31]. The estimated $r_s$ for our experiment (top axis of Fig. 4d) spans 13 – 26, showing that electrons in 2D TMDs are well into the strong interaction regime. In the inset of Fig. 4d, we show the doping-induced mass renormalization predicted by the result originally of a high-density calculation $m_r \approx m_{r0}(1 + 0.043 r_s)$ ($m_{r0} \approx 0.2 m_0$ [31]) that has been shown a fairly good approximation even well beyond the high-density limit [33]. Qualitative agreement between the experimental result and the simple model is seen. The mass increases monotonically with the interaction strength before doping into the spin-split upper conduction band c1. The step rise corresponds to doping into c1 and doubling the electron species. In conclusion, our experiment has demonstrated the unique LL structure in monolayer TMDs. The robust valley and spin splitting of the zero-energy LL in TMDs, resulted from inversion symmetry breaking and SOI, respectively, open up new possibilities for studies of unconventional LL physics and quantum Hall effect in a 2D semiconductor.

## Methods

### Device fabrication

Dual-gate field-effect transistors (FETs) of monolayer $WSe_2$ studied in this work have been built from individual van der Waals (vdW) layers using a mechanical transfer method developed by Wang et al. [34]. Thin flakes of hexagonal boron nitride (hBN), few-layer graphene and monolayer $WSe_2$ were first exfoliated from bulk crystals onto silicon substrates covered by a 300 nm thermal oxide layer. A thin layer of polypropylene carbonate (PPC) on polydimethylsiloxane (PDMS) supported by a glass slide was prepared as a stamp. The vdW flakes were picked up layer-by-layer by the stamp using micromanipulators under a microscope. The finished multilayer stack was then released onto a silicon substrate (with a 100 nm thermal oxide layer) with pre-patterned gold electrodes. The substrate was aligned such that each of the four graphene electrodes (source, drain, top gate and back gate) was in contact with only one gold electrode. An optical image of the finished device and its schematic side view are shown in Fig. 1a of the main text. The residual PPC on the device was dissolved in anisol before optical measurements. Monolayer $WSe_2$ has been fully encapsulated by hBN to achieve high quality required for the observation of inter-LL transitions. Three devices have been measured at low temperature and two of them showed clear LL structures below 9 Tesla.

### Optical spectroscopy at low temperature and under high magnetic fields

Optical spectroscopy with a sub-micron spatial resolution at 3 K and under a magnetic field up to 9 T was performed in an Attocube close-cycle cryostat (attoDry1000) with a confocal microscopy insert. Optical radiation was coupled into and out of the system using free-space optics. A high numerical aperture (N.A. = 0.8) objective was used to focus the excitation beam onto the device and also collect the reflection/emission from it. The collected radiation was detected by a spectrometer equipped with a charge-coupled-device (CCD). For the reflectance contrast measurements, broadband radiation from a supercontinuum light source was employed. A quarter-waveplate in the incident beam path was used to select the left and right circular polarization. The excitation power on the device was kept below 5 μW to minimize heating effects. The reflectance contrast spectrum $\frac{\Delta R}{R}$ was obtained by comparing the reflectance from the part of the device with the monolayer $WSe_2$ channel ($R'$) and without the channel ($R$) as $\frac{\Delta R}{R} \equiv \frac{R'-R}{R}$. For photoluminescence (PL) measurements, the excitation light source was a HeNe laser at 632.8 nm with an incident power < 100 μW on the device. A long-pass filter removes the laser line for detection of the PL spectrum by the spectrometer and CCD.

### Acknowledgement

The experimental research was supported by the National Science Foundation through Grant No. DMR-1410407, 1420451 (sample and device fabrication) and the Air Force Office of Scientific Research under grant FA9550-14-1-0268 (spectroscopy in high fields). Support for data analysis was provided by the US Department of Energy, Office of Basic Energy Sciences under Award No. DESC0012635 (J.S.) and the Air Force Office of Scientific Research under grant FA9550-16-1-0249 (K.F.M.).

**References:**

1. Xu, X., Yao, W., Xiao, D. & Heinz, T.F. Spin and pseudospins in layered transition metal dichalcogenides. *Nat Phys* **10**, 343-350 (2014).
2. Mak, K.F. & Shan, J. Photonics and optoelectronics of 2D semiconductor transition metal dichalcogenides. *Nat Photon* **10**, 216-226 (2016).
3. Cao, T. *et al.* Valley-selective circular dichroism of monolayer molybdenum disulphide. *Nat Commun* **3**, 887 (2012).
4. Zeng, H., Dai, J., Yao, W., Xiao, D. & Cui, X. Valley polarization in MoS2 monolayers by optical pumping. *Nat Nano* **7**, 490-493 (2012).
5. Mak, K.F., He, K., Shan, J. & Heinz, T.F. Control of valley polarization in monolayer MoS2 by optical helicity. *Nat Nano* **7**, 494-498 (2012).
6. Jones, A.M. *et al.* Optical generation of excitonic valley coherence in monolayer WSe2. *Nat Nano* **8**, 634-638 (2013).
7. Mak, K.F., McGill, K.L., Park, J. & McEuen, P.L. The valley Hall effect in MoS2 transistors. *Science* **344**, 1489-1492 (2014).
8. Lee, J., Mak, K.F. & Shan, J. Electrical control of the valley Hall effect in bilayer MoS2 transistors. *Nat Nano* **11**, 421-425 (2016).
9. Li, X., Zhang, F. & Niu, Q. Unconventional Quantum Hall Effect and Tunable Spin Hall Effect in Dirac Materials: Application to an Isolated MoS2 Trilayer. *Physical Review Letters* **110**, 066803 (2013).
10. Cai, T.Y. *et al.* Magnetic control of the valley degree of freedom of massive Dirac fermions with application to transition metal dichalcogenides. *Physical Review B* **88**, 115140 (2013).
11. Cui, X. *et al.* Multi-terminal transport measurements of MoS2 using a van der Waals heterostructure device platform. *Nat Nano* **10**, 534-540 (2015).
12. R. E. Prange & Girvin, S.M. *The quantum Hall effect* (eds. R. E. Prange & Girvin, S.M.) (Springer-Verlag New York, 1990).
13. Gunawan, O. *et al.* Valley susceptibility of an interacting two-dimensional electron system. *Physical Review Letters* **97**, 186404 (2006).
14. Shkolnikov, Y.P., De Poortere, E.P., Tutuc, E. & Shayegan, M. Valley splitting of AlAs two-dimensional electrons in a perpendicular magnetic field. *Physical Review Letters* **89**, 226805 (2002).
15. Castro Neto, A.H., Guinea, F., Peres, N.M.R., Novoselov, K.S. & Geim, A.K. The electronic properties of graphene. *Reviews of Modern Physics* **81**, 109-162 (2009).
16. Lensky, Y.D., Song, J.C.W., Samutpraphoot, P. & Levitov, L.S. Topological Valley Currents in Gapped Dirac Materials. *Physical Review Letters* **114**, 256601 (2015).
17. Weitz, R.T., Allen, M.T., Feldman, B.E., Martin, J. & Yacoby, A. Broken-Symmetry States in Doubly Gated Suspended Bilayer Graphene. *Science* **330**, 812-816 (2010).
18. Rutter, G.M. *et al.* Microscopic polarization in bilayer graphene. *Nature Physics* **7**, 649-655 (2011).
19. Zhao, Y., Cadden-Zimansky, P., Jiang, Z. & Kim, P. Symmetry Breaking in the Zero-Energy Landau Level in Bilayer Graphene. *Physical Review Letters* **104**, 066801 (2010).
20. Nomura, K. & MacDonald, A.H. Quantum Hall ferromagnetism in graphene. *Physical Review Letters* **96**, 256602 (2006).
21. Young, A.F. *et al.* Spin and valley quantum Hall ferromagnetism in graphene. *Nature Physics* **8**, 550-556 (2012).

22. Duggan, G. Theory of heavy-hole magnetoexcitons in GaAs-(Al,Ga)As quantum-well heterostructures *Physical Review B* **37**, 2759-2762 (1988).
23. Fallahazad, B. *et al.* Shubnikov-de Haas Oscillations of High-Mobility Holes in Monolayer and Bilayer WSe2: Landau Level Degeneracy, Effective Mass, and Negative Compressibility. *Physical Review Letters* **116**, 086601 (2016).
24. Rose, F., Goerbig, M.O. & Piechon, F. Spin- and valley-dependent magneto-optical properties of MoS2. *Physical Review B* **88**, 125438 (2013).
25. Chu, R.-L. *et al.* Valley-splitting and valley-dependent inter-Landau-level optical transitions in monolayer MoS2 quantum Hall systems. *Physical Review B* **90**, 045427 (2014).
26. Li, Y.L. *et al.* Valley Splitting and Polarization by the Zeeman Effect in Monolayer MoSe2. *Physical Review Letters* **113**, 266804 (2014).
27. MacNeill, D. *et al.* Breaking of Valley Degeneracy by Magnetic Field in Monolayer MoSe2. *Physical Review Letters* **114**, 037401 (2015).
28. Aivazian, G. *et al.* Magnetic control of valley pseudospin in monolayer WSe2. *Nat Phys* **11**, 148-152 (2015).
29. Srivastava, A. *et al.* Valley Zeeman effect in elementary optical excitations of monolayer WSe2. *Nat Phys* **11**, 141-147 (2015).
30. Wang, G. *et al.* Magneto-optics in transition metal diselenide monolayers. *2D Materials* **2**, 034002 (2015).
31. Liu, G.-B., Shan, W.-Y., Yao, Y., Yao, W. & Xiao, D. Three-band tight-binding model for monolayers of group-VIB transition metal dichalcogenides. *Physical Review B* **88**, 085433 (2013).
32. Stern, M., Garmider, V., Umansky, V. & Bar-Joseph, I. Mott transition of excitons in coupled quantum wells. *Physical Review Letters* **100**, 256402 (2008).
33. Krakovsky, A. & Percus, J.K. Quasiparticle effective mass for the two- and three-dimensional electron gas. *Physical Review B* **53**, 7352-7356 (1996).
34. Wang, L. *et al.* One-Dimensional Electrical Contact to a Two-Dimensional Material. *Science* **342**, 614-617 (2013).

**Figures and figure captions:**

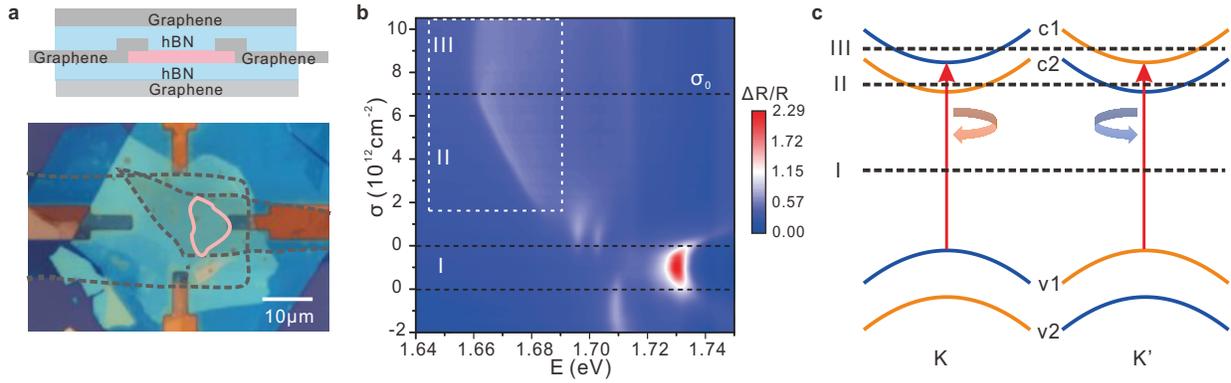

**Figure 1| Optical transitions in dual-gated monolayer WSe$_2$. a**, Optical microscope image and schematic side view of a dual-gate field-effect transistor of monolayer WSe$_2$ with graphene contact and gate electrodes. WSe$_2$ is fully encapsulated by hBN. The boundaries of the WSe$_2$ channel and the graphene gates are shown in pink solid line and grey dashed lines, respectively. **b**, The reflectance contrast contour under $B = 0$ shows three doping regime I, II and III. Doping density $\sigma_0$ corresponds to the Fermi energy at the bottom of band c1. The dashed box is the region of interest. **c**, Electronic band structure of monolayer WSe$_2$ at the K and K' valley and the valley contrasting optical selection rules for the A exciton features shown in **b**. (c1, c2) and (v1, v2) are the spin-split conduction and valence bands. Optical transitions are allowed between bands of the same electron spin (shown in the same color). The dotted lines show three representative Fermi energies from doping regime I, II and III, respectively.

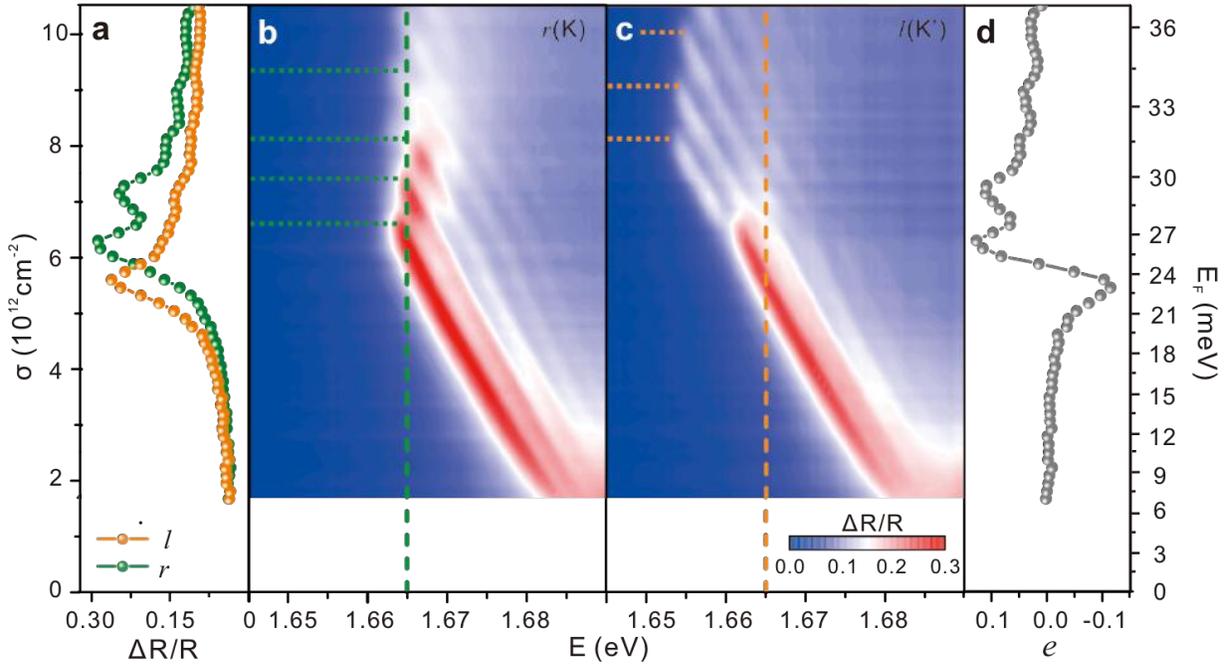

**Figure 2| Valley contrasting inter-LL transitions. a**, Reflectance contrast of monolayer $WSe_2$ at a representative photon energy of 1.665 eV as a function of doping density $\sigma$ for K (*r* channel) and K' (*l* channel) valley. **b, c**, Reflectance contrast contour as a function of doping density $\sigma$ (left axis), Fermi energy $E_F$ (right axis), and photon energy $E$ (bottom axis) under 9 T for K and K' valleys. The doping densities $\sigma_n^*$ required to fill each LL are marked by horizontal dotted lines. **d**, Optical ellipticity as a function of doping or Fermi energy calculated from **a**.

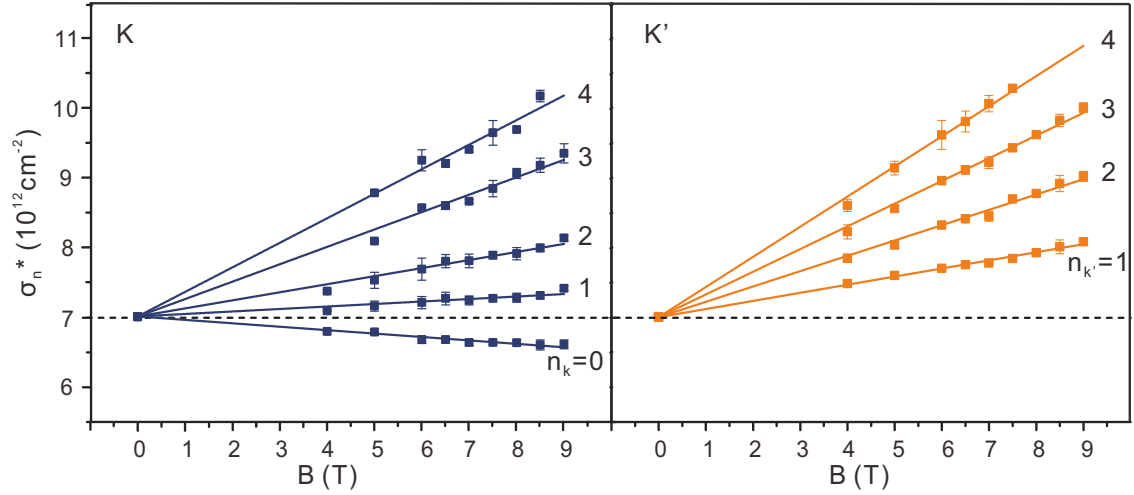

**Figure 3| Magnetic-field dispersion of valley-spin polarized LLs.** The doping densities $\sigma_n^*$ required to fill LL $n_K = 0 - 4$ (at the K valley) and $n_{K'} = 1 - 4$ (at the K' valley) are shown as a function of magnetic field $B$. Solid lines are linear fits to the experimental data (symbols). The error bar is the uncertainty in determining $\sigma_n^*$ from Fig. 2.

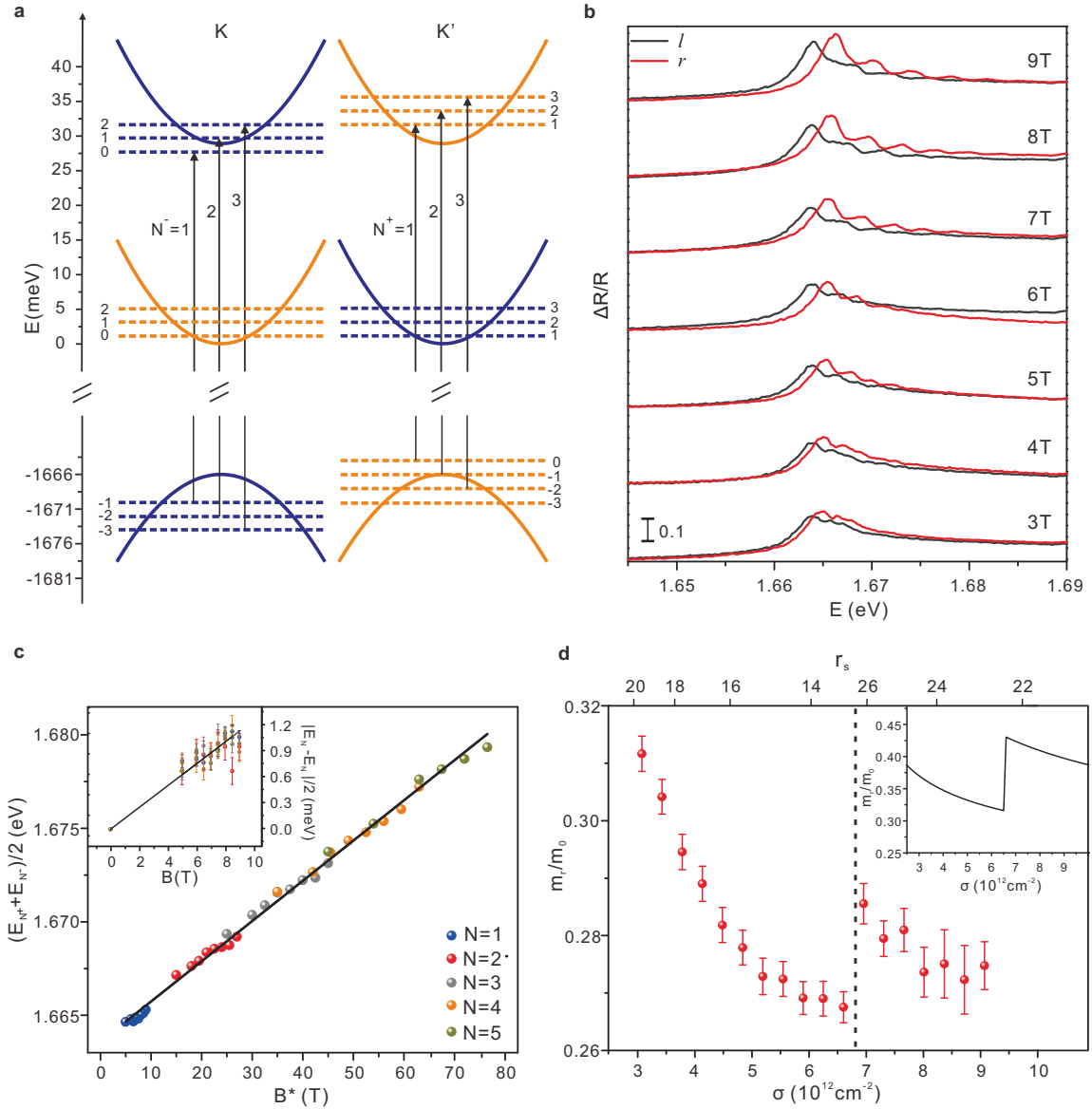

**Figure 4| LL structure of monolayer WSe$_2$. a**, The LL structure of monolayer WSe$_2$ at 9 T for doping at $5.9\times10^{12}$ cm$^{-2}$ constructed from the results of Fig. 2 & 3. Band v1, c1 and c2 are shown in solid lines with bands of the same electron spin in the same color. The LLs are represented by dashed lines. Black lines with an arrowhead show the valley and spin contrasting optical selection rules. **b**, The left- (*l*) and right-handed (*r*) reflectance contrast spectra at a fixed doping density of $\sigma = 5.9\times10^{12}$cm$^{-2}$ under differing magnetic fields (displaced vertically for clarity). The vertical bar shows the scale for contrast equal to 0.1. **c**, The valley averaged inter-LL transition energies $(E_{N^+} + E_{N^-})/2$ for the five lowest energy transitions ($N$ = 1-5) as a function of reduced magnetic field $B^* = (2N - 1)B$. The inset shows $|E_{N^+} - E_{N^-}|/2$ as a function of $B$. Solid lines are linear fits to the experimental data (symbols). **d**, The reduced mass $m_r$, determined from the slope of **c**, as a function of doping density. The error bar is the

uncertainty propagated from the standard deviation of the mean for the slope from the least squares fit. Values of $r_s$ are shown in the top axis. The inset shows the dependence of $m_r \approx m_{r0}(1 + 0.043 r_s)$ on the experimental $r_s$ value.